\documentclass[a4paper,10pt,reqno]{amsart}
\usepackage{a4wide}
\usepackage{amsfonts,amsmath,amssymb}
\usepackage{graphicx}
\usepackage{tikz}

\theoremstyle{remark}

\def\be{\begin{equation}}
\def\ee{\end{equation}}
\def\bea{\begin{eqnarray}}
\def\eea{\end{eqnarray}}

\def\rme{\mathrm{e}}
\def\rmi{\mathrm{i}}
\def\scq{{\textsc q}}

\begin{document}

\thispagestyle{plain}

\title{Hofstadter point spectrum trace  and the almost Mathieu operator}

\author{St\'ephane OUVRY (*), Stephan WAGNER (**) and Shuang WU (*)}

\date{\today}

\begin{abstract}
We consider  point spectrum traces in the Hofstadter model. We show how  to recover the full quantum Hofstadter trace by integrating  these point spectrum traces with the appropriate free density of states on the lattice. This construction is then generalized to the almost Mathieu operator and its $n$-th moments which can be  expressed in terms of generalized Kreft coefficients.
\end{abstract}

\maketitle
(*) LPTMS, CNRS-Facult\'e des Sciences d'Orsay, Universit\'e Paris Sud, 91405 Orsay Cedex, France

(**)  Department of Mathematical Sciences, Stellenbosch University, Matieland 7602, South Africa

\section{Introduction }
In \cite{nous} we focused on the algebraic area  generating function  of closed lattice walks of a given length $n$ ($n$ is then necessarily even) 
\[ Z_n(e^{i\gamma})\] 
evaluated at $e^{i\gamma}$, a root of unity.
  One reason for studying this  quantity  arises from  the connection  of the algebraic area distribution of random curves  to the quantum spectrum of a charged particle in a perpendicular magnetic field.
 In the lattice case  at hand,  the mapping is on the quantum Hofstadter model \cite{Hofstadter} of a particle hopping on a two-dimensional lattice  in  a  magnetic flux $\gamma$, counted in unit of the flux quantum.   More precisely  $Z_n(e^{i\gamma})$ is mapped  on the $n$-th moment ${\rm Tr}\:H_{\gamma}^{n}$ of the Hofstadter Hamiltonian $H_{\gamma}$  --thereafter referred to as the quantum Hofstadter trace--
\be \nonumber Z_{n}(e^{i\gamma})={\rm Tr}\:H_{\gamma}^{n}\ee
by virtue of which evaluating  $Z_n(e^{i\gamma})$  for classical lattice walks gives an expression  \cite{nous} for the  Hofstadter quantum trace ${\rm Tr}\:H_{\gamma}^{n}$.  In the simplified case of a rational flux, not surprisingly, the trace ${\rm Tr}\:H_{\gamma}^{n}$ can  be written in terms of the Kreft coefficients \cite{Kreft} which    encode   the    Schr\"odinger equation for the Hofstadter model
 \be \Phi_{m+1}+\Phi_{m-1}+2\cos(k_y+\gamma m)\Phi_{m}=E\Phi_{m}.\label{eq}\ee 
 
  One would like to generalize this construction to  the almost Mathieu operator case
 \be \Phi_{m+1}+\Phi_{m-1}+\lambda\cos(k_y+\gamma m)\Phi_{m}=E\Phi_{m}\label{eqMathieu}\ee 
 where $\lambda$ is now a  free parameter.  
This operator, among other things,  plays an important role in the characterization of the fractal structure of the Hofstadter spectrum \cite{Last}. Physically, it describes  a quantum particle hopping on a lattice with horizontal and vertical amplitudes in a ratio $\lambda$.

We will   first rederive the results of \cite{nous} by starting from  point spectrum traces --to be defined later--  and  integrating them with the appropriate free density of states on the lattice in order to recover  the  quantum trace ${\rm Tr}\:H_{\gamma}^{n}$. This approach is original and gives a new light on the results obtained in \cite{nous}. The generalization to the almost Mathieu case will then follow provided that one can extend accordingly the Kreft   coefficients  construction to the  $\lambda\ne 2$ case. Finally we will discuss in the conclusion some direct links which can be established with  current activities in the field \cite{Hatsuda}.

\section{A reminder}

In the commensurate case  with a rational flux
$
\gamma  = 2 \pi {p}/{q}
$, with $p$ and $q$ co-primes,  the lattice Hofstadter eigenstates $\psi_{m,n}= e^{i n k_y}\Phi_m$ 
are $q$-periodic  on the horizontal axis
$\Phi_{m + q} = e^{i q k_x } \Phi_m$.  The Schr\"odinger equation \eqref{eq} then reduces  to  a $q\times q$ secular matrix
$m_{p/q}(E,k_x,k_y)$  acting with zero output on the $q$-components eigenvector $\{ \Phi_0, \Phi_1,\ldots, \Phi_{q-1}\}$
\be
\label{matrixHof}
\begin{pmatrix}
 2 \cos ({k_y})-E& 1 & 0 & \cdots & 0 & e^{-i {q k_x}} \\
1 & 2 \cos ({k_y}+\frac{2\pi p}{q})-E & 1 & \cdots & 0 & 0 \\
0 & 1 & () & \cdots & 0 & 0 \\
\vdots & \vdots & \vdots & \ddots & \vdots & \vdots \\
0 & 0 & 0  & \cdots & () & 1 \\
e^{i {q k_x}}  & 0 & 0  & \cdots & 1 & 2 \cos ({k_y}+(q-1)\frac{2\pi p}{q})-E  \\
\end{pmatrix}  \begin{pmatrix} \Phi_0 \\ \Phi_1 \\ \Phi_2 \\ \vdots \\ \Phi_{q-2} \\ \Phi_{q-1} \end{pmatrix} = \begin{pmatrix}  0\\ 0 \\ 0 \\ \vdots \\ 0 \\ 0 \end{pmatrix}.\ee
The $q$ eigenenergies $E_1(k_x,k_y), E_2(k_x,k_y), \ldots, E_q(k_x,k_y)$  are the  roots of   $\det(m_{p/q}(E,k_x,k_y))=0$, which, thanks to the identity  
\be\label{Chambers} \det(m_{p/q}(E,k_x,k_y))=\det(m_{p/q}(E,0,0))-2 (-1)^q (\cos(q k_x)-1 + \cos(q k_y)-1),\ee
   rewrites (see \cite{Chambers}) as
\begin{equation} \det(m_{p/q}(E,0,0))=2 (-1)^q (\cos(q k_x)-1 + \cos(q k_y)-1).\label{eigen}
\end{equation}
The polynomial $b_{p/ q}(z)$  with coefficients $-a_{p/q}(2j)$
\be \label{poly} b_{p/ q}(z):=-\sum_{j=0}^{[\frac{q}{2}]} a_{p/q}(2j)z^{2j}\ee
($a_{p/q}(0)=-1$) materializes in  $\det(m_{p/q}(E,0,0))$  as 
\be\label{sososimple}
\det(m_{p/q}(E,0,0))+4(-1)^q=(-1)^q E^qb_{p/q}(1/E),
\ee
so that
\eqref{eigen} becomes
\begin{equation}\label{sosimple}
E^qb_{p/q}(1/E)=2(\cos(q k_x)+\cos(q k_y)).
\end{equation} 
The $a_{p/ q}(2j)$'s in $b_{p/q}(z)$ in (\ref{poly}) are  related  to the Kreft  coefficients  ${\rm  c}_{p/q}(2j)$  
\be \nonumber\det(m_{p/q}(E,0,0))+4(-1)^q=\sum_{j={q\over 2}-\lfloor{q\over 2}\rfloor}^{{q\over 2}}{\rm  c}_{p/q}(2j)E^{2j},\ee
so that $a_{p/q}(2j)={\rm  c}_{p/q}(q-2j)(-1)^{q+1}$.  One gets
\begin{equation}\label{thea}\small{
a_{p/q}(2j)=(-1)^{j+1}\sum_{k_1=0}^{q-2j}\sum_{k_2=0}^{k_1}\ldots\sum_{k_{j}=0}^{k_{j-1}} 4\sin ^2\left(\frac{\pi  (k_1+2j-1) p}{q}\right)4\sin ^2\left(\frac{\pi  (k_2+2j-3) p}{q}\right)\ldots 4\sin ^2\left(\frac{\pi  (k_{j}+1) p}{q}\right)},
\end{equation}
with for building blocks 
\be\nonumber 4\sin ^2\left(\frac{\pi  (k+1) p}{q}\right)=e^{-ik_y}(1-e^{2 i \pi (k+1) p \over q})e^{ik_y}(1-e^{-{2 i \pi (k+1) p \over q}})=\alpha_{p/q}(k)\overline{\alpha}_{p/q}(k)\ee
where
\be\nonumber\alpha_{p/q}(k)=e^{-ik_y}(1-e^{2 i \pi (k+1) p \over q})\ee 
and $\overline{\alpha}_{p/q}(k)$ is the complex conjugate.
 How to  derive \eqref{thea} is explained in Kreft's paper \cite{Kreft} (see the appendix for details).  Note that  $a_{p/q}(2j)=0$ as soon as $q<2j$.

\noindent A closed expression for the quantum Hofstadter trace, which is defined as    
 \be\label{trace}
{\rm Tr}\; H_{2\pi p/ q}^n = \frac{1}{q}\int_{-\pi}^{\pi}\int_{-\pi}^{\pi} \frac{d k_x}{2\pi}\frac{d k_y}{2\pi} \sum_{r= 1}^{q} E_r^n(k_x,k_y), 
\ee 
  where one has integrated over the quasi-momenta  $k_x$ and $k_y\in[-\pi,\pi]$ the  sum  of  the $q$ eigenenergies $E_r(k_x,k_y)$---the roots of \eqref{sosimple}---at a power $n$,   has been obtained   \cite{nous} in terms of the Kreft coefficients
\begin{align}\label{sumformula}
{\rm Tr} H_{2\pi p/q}^{n} =& \frac{n}{q} \sum_{k \geq 0} \sum_{\substack{\ell_1,\ell_2,\ldots,\ell_{\lfloor q/2 \rfloor} \geq 0 \\ \ell_1 + 2\ell_2 + \cdots + \lfloor q/2 \rfloor \ell_{\lfloor q/2 \rfloor} = n/2 - qk}}
\frac{\binom{\ell_1+\ell_2 + \cdots + \ell_{\lfloor q/2 \rfloor} + 2k}{\ell_1,\ell_2,\ldots,\ell_{\lfloor q/2 \rfloor},2k}}{\ell_1+\ell_2 + \cdots + \ell_{\lfloor q/2 \rfloor} + 2k} \binom{2k}{k}^2\\\nonumber&  \prod_{j = 1}^{\lfloor q/2 \rfloor} a_{p/q}(2j)^{\ell_j},
\end{align}
with the generating function 
\begin{equation}\label{eq:K-eq}
\sum_{n \geq 0} {\rm Tr} H_{2\pi p/q}^{n} z^n = \Big( 1 - \frac{zb'_{p/ q}(z)}{qb_{p/ q}(z)}\Big) \frac{2}{\pi} K \Big( \frac{16z^{2q}}{b_{p/ q}(z)^2} \Big),
\end{equation}
where $K$ is the complete elliptic integral of the first kind $\frac{2}{\pi} K(16x)=\sum_{k \geq 0} \binom{2k}{k}^2 x^k$.

\section{ Point spectrum trace formula   and density of states}
\subsection{Mid-band trace formula:} One aims at generalizing \eqref{sumformula} and \eqref{eq:K-eq} to the almost Mathieu operator case  \eqref{eqMathieu}. 
To achieve this goal first of all one  remarks that the Hofstadter trace  \eqref{sumformula},  valid for all $n$    and $q$,  coincides when  $q>n/2$  with the mid-band traces  
given     for the particular cases  $n=2,\ldots,10$ in \cite{Lipan}. The mid-band trace 
\be\nonumber {\rm Tr_{0}} {H^n_{2\pi p/q}}=\frac{1}{q} \sum_{r= 1}^{q} E_r^n(0),\ee
to be distinguished from the  quantum Hofstadter trace \eqref{trace}, are taken solely on the mid-band energies\footnote{The mid-band energies have been extensively studied  and  \cite{Wiegmann} are attainable  from  Bethe-ansatz equations for the quantum group $U_q(sl_2)$.} $E_r(0)$, the roots of
\begin{equation}\label{yes}
E^q b_{p/ q}(1/E)=0 
 \end{equation} 
 to be distinguished from the $E_r(k_x,k_y)$'s,  the roots of  \eqref{sosimple} 
 (for an illustration see Figures 1  and 2).
 \begin{figure}
\includegraphics[scale=.5]{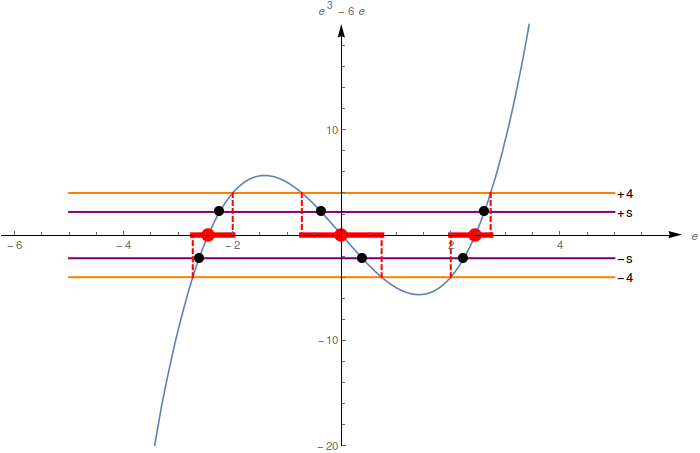}\label{fig}
\caption{$p=1$, $q=3$: $E^qb_{p/q}(1/E)=E^3-6E$, the 3 horizontal red segments are the energy bands; the 3 red dots are the mid-band energies; the 6 black dots are the $\pm s$ energies.}
\centering
\end{figure}
 \begin{figure}
\centering
\includegraphics[scale=.5]{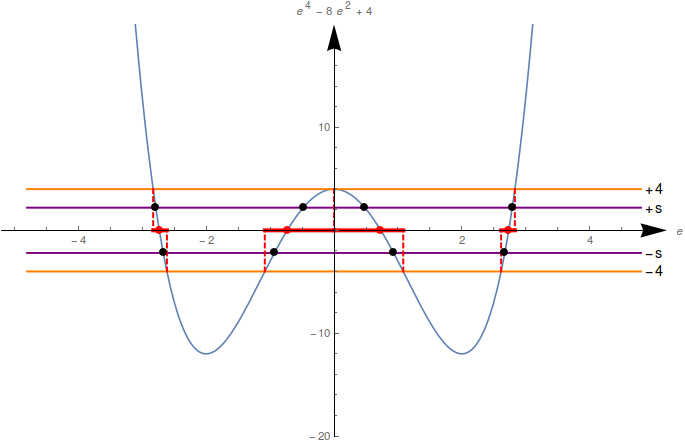}\label{figurebis}
\caption{$p=1$, $q=4$: $E^qb_{p/q}(1/E)=E^4-8E^2+4$, the 4 horizontal red segments are the energy bands; the 4 red dots are the mid-band energies; the 8 black dots are the $\pm s$ energies.}
\centering
\end{figure}
 It is quite straightforward to  obtain  for all $n$   and $q$    the  mid-band trace  formula
\be
{\rm Tr_{0}} H_{2\pi p/q}^{n} = \frac{n}{q} \sum_{\substack{\ell_1,\ell_2,\ldots,\ell_{\lfloor q/2 \rfloor} \geq 0 \\ \ell_1 + 2\ell_2 + \cdots + \lfloor q/2 \rfloor \ell_{\lfloor q/2 \rfloor} = n/2}}
\frac{ \binom{\ell_1+\ell_2 + \cdots + \ell_{\lfloor q/2 \rfloor} }{\ell_1,\ell_2,\ldots,\ell_{\lfloor q/2 \rfloor}}}{\ell_1+\ell_2 + \cdots + \ell_{\lfloor q/2 \rfloor} } \prod_{j = 1}^{\lfloor q/2 \rfloor} a_{p/q}(2j)^{\ell_j}.\label{sumformulapoint}
\ee
 One can check   that  the Hofstadter trace \eqref{sumformula} and the  mid-band  trace  \eqref{sumformulapoint} indeed coincide 
provided that  $q>n/2$ since  then the summation index  $k$ in \eqref{sumformula}  necessarily vanishes.  A   qualitative interpretation for this fact can stem from the classical picture of lattice walks contained in the quantum periodic cell of horizontal length $q$ (the  lattice  walk the farthest on the horizontal axis from  the origin  indeed goes to  a distance  $ n/2$). 

 The mid-band trace 
  \eqref{sumformulapoint}  has a  simple    combinatorial interpretation: it
  is the sum of  products  
 of the $a_{p/q}(2j)^{\ell_j}$'s
corresponding to  partitions of $n$ in  even integer parts of  size $2j$ (no larger than $q$ since  $a_{p/q}(2j)=0$ as soon as $q<2j$), i.e.   
   $2\ell_1 + 4\ell_2 + \cdots + 2\lfloor q/2 \rfloor \ell_{\lfloor q/2 \rfloor} = n$,
multiplied by  the multinomial  weight.

The generating function    follows as
\begin{equation}\label{eq:K-eqmid}
\sum_{n \geq 0,n\;{\rm even}} {\rm Tr_{0}} {H^n_{2\pi p/q}} z^n = 1 - \frac{zb'_{p/ q}(z)}{qb_{p/ q}(z)},
\end{equation}
which amounts to nothing else but \eqref{yes}, i.e. the $E_r(0)$'s are the roots of  $E^q b_{b/q}(1/E)=0$. To see this rewrite
\begin{align*}
\sum_{n \geq 0, n\;{\rm even}} {\rm Tr_{0}} {H^n_{2\pi p/q}} z^n &= \sum_{n \geq 0, n\;{\rm even}}{1\over q} \Big( \sum_{r=1}^q E_r^n(0) \Big) z^n ={1\over q}\sum_{r=1}^q\sum_{n \geq 0, n\;{\rm even}} ( zE_r(0))^n \\  &={1\over q}\sum_{r=1}^q{1\over 1-z^2E_r^2(0)}={1\over 2q}\sum_{r=1}^q \Big({1\over 1-zE_r(0)}+{1\over 1+zE_r(0)}\Big)={1\over q}\sum_{r=1}^q{1\over 1-zE_r(0)},\nonumber
\end{align*}
so that \eqref{eq:K-eqmid} becomes
\begin{equation}\label{eq:K-eqmidbis}
{1\over q}\sum_{r=1}^q{1\over 1-z E_r(0)}= 1 - \frac{zb'_{p/ q}(z)}{qb_{p/ q}(z)}.
\end{equation}  
On the other hand \eqref{yes}  rewrites as
\be \nonumber ({1/ z})^q b_{p/q}(z)=\prod_{r=1}^q({1/ z}-E_r(0)), \ee
that is 
\be \nonumber  b_{p/q}(z)=\prod_{r=1}^q({1}-z E_r(0)), \ee
so that 
\be \nonumber  \big(\log(b_{p/q}(z))\big)'=\sum_{r=1}^q{-E_r(0)\over {1}-z E_r(0)}= {1\over z}\sum_{r=1}^q \Big(1-{1\over 1-z E_r(0)}\Big)\ee
or
\be \nonumber  {b_{p/q}(z)'\over b_{p/q}(z)}= {1\over z} \Big(q-\sum_{r=1}^q{1\over 1-z E_r(0)} \Big),\ee
which is  \eqref{eq:K-eqmidbis}.
\subsection{$\pm s$ trace formula:}
More generally one  can  consider the $\pm s$  spectrum traces 
\be \nonumber {\rm Tr_{\pm s}} {H^n_{2\pi p/q}} ={1\over 2q}\Big(\sum_{r=1}^q E_r^n(s)+\sum_{r=1}^q E_r^n(-s)\Big)\ee 
taken on the $2q$ roots $E_r(s)$ and $E_r(-s)$ of
\be\label{sss} E^q b_{p/ q}(1/E)= s\;{\rm and}\;E^q b_{p/ q}(1/E)=- s\ee 
with $-4\le s\le 4$ since  $-4\le 2(\cos(q k_x)+\cos(q k_y))\le 4$ in \eqref{sosimple}. The mid-band spectrum  is obtained for $s=0$, and the edge-band spectrum for $s=4$,  of particular interest as well (see Figures 1 and 2). 

Following the same line of reasoning as for the mid-band spectrum, one obtains  the $\pm s$  spectrum traces  generating function   
\begin{align}
\sum_{n \geq 0,n\;{\rm even}} {\rm Tr_{\pm s}} {H^n_{2\pi p/q}}  z^n &= 1 - \frac{z(b_{p/ q}(z)-sz^q)'}{2q(b_{p/ q}(z)-sz^q)}- \frac{z(b_{p/ q}(z)+sz^q)'}{2q(b_{p/ q}(z)+sz^q)} \nonumber\\ 
&=1 - \frac{z(b^2_{p/ q}(z)-s^2z^{2q})'}{2q(b^2_{p/ q}(z)-s^2z^{2q})}\nonumber\\&=\Big( 1 - \frac{zb'_{p/ q}(z)}{qb_{p/ q}(z)}\Big){1\over 1-s^2(\frac{z^{q}}{b_{p/ q}(z)})^2}, \label{eq:K-eqs}
\end{align}
which rightly reduces to the mid-band case \eqref{eq:K-eqmid} when $s=0$.
 Again \eqref{eq:K-eqs} is a mere rewriting of the $E_r(s)$'s and $E_r(-s)$'s being the roots of $E^q b_{p/ q}(1/E)=\pm s$  in \eqref{sss}.

One  has likewise  the $\pm s$  trace  formula
\be
{\rm Tr_{\pm s}} H_{2\pi p/q}^{n} = \frac{n}{q} \sum_{k \geq 0} \sum_{\substack{\ell_1,\ell_2,\ldots,\ell_{\lfloor q/2 \rfloor} \geq 0 \\ \ell_1 + 2\ell_2 + \cdots + \lfloor q/2 \rfloor \ell_{\lfloor q/2 \rfloor} = n/2 - qk}}
\frac{ \binom{\ell_1+\ell_2 + \cdots + \ell_{\lfloor q/2 \rfloor} + 2k}{\ell_1,\ell_2,\ldots,\ell_{\lfloor q/2 \rfloor},2k}}{\ell_1+\ell_2 + \cdots + \ell_{\lfloor q/2 \rfloor} + 2k} s^{2k}\prod_{j = 1}^{\lfloor q/2 \rfloor} a_{p/q}(2j)^{\ell_j}\label{sumformulas}
\ee
which again reduces to the mid-band trace \eqref{sumformulapoint} when $s=0$. The combinatorial interpretation  of (\ref{sumformulas}) is again simple  in terms of
      products of $s^2$ and of the $a_{p/q}(2j)$'s 
corresponding to all partitions of $n$ in even integer parts   of size   $2q$ and  $2j$ respectively.

 Again,  as soon as $q>n/2$, the trace ${\rm Tr_{\pm s}} {H^n_{2\pi p/q}}$ in \eqref{sumformulas} does not depend on $s$ anymore and coincides with the  Hofstadter trace \eqref{sumformula}  and, by the same token, with the   mid-band trace  \eqref{sumformulapoint}.

\subsection{Density of states:}
One remarks next that  the Hofstadter quantum trace  ${\rm Tr} {H^n_{2\pi p/q}}$ in  \eqref{sumformula}  
can be  recovered from the $\pm s$ spectrum trace ${\rm Tr_{\pm s}} {H^n_{2\pi p/q}}$ in \eqref{sumformulas} if one replaces    $s^{2k}$  by ${2k\choose k}^2$. This replacement amounts to an integration  of \eqref{sumformulas} over   $s\in[-4,4]$ with a density of state $\rho(s)$ 
\be \int_{-4}^{4}{\rm Tr_{\pm s}} {H^n_{2\pi p/q}} \rho(s) ds= {\rm Tr} {H^n_{2\pi p/q}}  \label{integration} \ee
such that        
\be \int_{-4}^{4}s^{2k} \rho(s) ds={2k\choose k}^2 
\label{required}\ee 
Clearly  the free density of states on the 2d lattice,
\be\label{itis}\rho(s)=1/(2\pi^2) K(1-s^2/16),\ee    enforces \eqref{required}: it is the  density of states  for the  spectrum  $s=2(\cos(q k_x)+\cos(q k_y))$.
Again, as soon as $q>n/2$, the integration in \eqref{integration} becomes trivial, i.e. the identity\footnote{In the spirit of \cite{Wannier}, one can     make   the change of variable  $s=\pm E^qb_{p/q}(1/E)$ in \eqref{required} to  obtain
\be\nonumber
\int_{-4}^4(E^qb_{p/q}(1/E))^{2k}\rho_{p/q}(E) dE={2k\choose k}^2\ee 
where $\rho_{p/q}(E)$ is the Hofstadter density of states.
  This yields the trace sum rule
\be\nonumber
{\rm Tr}(H^q_{2\pi p/q}b_{p/q}(1/H_{2\pi p/q}))^{2k}={2k\choose k}^2\ee 
  or
\be\nonumber
{\rm Tr} \big(\sum_{j=0}^{[{q\over 2}]}a_{p/ q}(2j)H^{q-2j}_{2\pi p/q} \big)^{2k}={2k\choose k}^2,\ee 
which  is an inversion of \eqref{sumformula}.}.

 For the generating function as well, which is given by
\be \nonumber \int_{-4}^{4}\sum_n {\rm Tr_{\pm s}} {H^n_{2\pi p/q}} z^n \rho(s) ds= \sum_n {\rm Tr} {H^n_{2\pi p/q}} z^n,  \ee
 one uses \eqref{eq:K-eq}, \eqref{eq:K-eqs}  and \eqref{itis} so that
\be \nonumber \int_{-4}^{4} \Big(1 - \frac{z (b^2_{p/ q}(z)-s^2z^{2q})'}{2q(b^2_{p/ q}(z)-s^2z^{2q})} \Big) {1\over 2\pi^2 }K\left(1-{s^2\over 16}\right) ds= \Big( 1 - \frac{zb'_{p/ q}(z)}{qb_{p/ q}(z)}\Big) \frac{2}{\pi} K \Big(16 \Big(\frac{z^{q}}{b_{p/ q}(z)} \Big)^2 \Big)\ee
has to be satisfied.
After simplification this narrows down  to 
\be \nonumber\int_{-4}^{4}  {1\over 2\pi^2 }K\left(1-{s^2\over 16}\right){1\over 1-s^2(\frac{z^{q}}{b_{p/ q}(z)})^2} ds=  \frac{2}{\pi} K \Big(16 \Big(\frac{z^{q}}{b_{p/ q}(z)}\Big)^2 \Big),\ee
i.e. 
\be
\nonumber \int_{-4}^4 K\left(1-{s^2\over 16}\right){1\over 1-s^2 z^2}ds=4\pi {K(16 z^2)},\ee
which  is nothing but 
the complete elliptic integral of the first kind being  the generating function for the square of the binomial coefficients, i.e.  ${2\over \pi}K(16 z^2)=\sum_{k} {2k\choose k}^2\ z^{2k}$.

\section{$\lambda\ne 2$ :  point spectrum trace formula   and density of states}

All these considerations can be extended when $\lambda\ne 2$ to the almost Mathieu case \eqref{eqMathieu}  
whose   secular matrix  is 
\be\label{matrixHoflambda}
 m_{p/q}^{(\lambda)}(E,k_x,k_y)=\begin{pmatrix}
 \lambda \cos ({k_y})-E& 1 & 0 & \cdots & 0 & e^{-i {q k_x}} \\
1 & \lambda \cos ({k_y}+\frac{2\pi p}{q})-E & 1 & \cdots & 0 & 0 \\
0 & 1 & () & \cdots & 0 & 0 \\
\vdots & \vdots & \vdots & \ddots & \vdots & \vdots \\
0 & 0 & 0  & \cdots & () & 1 \\
e^{i {q k_x}}  & 0 & 0  & \cdots & 1 & \lambda \cos ({k_y}+(q-1)\frac{2\pi p}{q})-E \end{pmatrix}.
\ee
Thanks to the identity 
\be \nonumber\det(m_{p/q}^{(\lambda)}(E,k_x,k_y))=\det(m_{p/q}^{(\lambda)}(E,0,0))-2 (-1)^q (\cos(q k_x)-1 +({\lambda/ 2})^q (\cos(q k_y)-1)),\ee
the Schr\"odinger equation becomes
\begin{equation} \det(m_{p/q}^{(\lambda)}(E,0,0))=2 (-1)^q (\cos(q k_x)-1 + ({\lambda/ 2})^q(\cos(q k_y)-1)).
\label{eigenlambda}
\end{equation}
Again one introduces the polynomial
 $b_{p/ q}^{(\lambda)}(z)$  and its coefficients $-a_{p/q}^{(\lambda)}(2j)$ (with $a_{p/q}^{(\lambda)}(0)=-1$)
\be \label{b_lambda} b_{p/ q}^{(\lambda)}(z)=-\sum_{j=0}^{[{q\over 2}]} a_{p/q}^{(\lambda)}(2j)z^{2j}\ee
 such that $\det(m_{p/q}^{(\lambda)}(E,0,0))$   rewrites as
\be\label{surelambda}\det(m_{p/q}^{(\lambda)}(E,0,0))+2(-1)^q(1+({\lambda/ 2})^q)=(-1)^q E^qb_{p/q}^{(\lambda)}(1/E),
\ee
so that \eqref{eigenlambda} becomes
\be\label{good} E^qb_{p/q}^{(\lambda)}(1/E)=2(\cos(q k_x)+({\lambda/ 2})^q \cos(q k_y)).\ee
In the appendix we show how to get a closed expression for the generalized Kreft coefficients $a_{p/ q}^{(\lambda)}(2i)$'s in \eqref{b_lambda}  following the steps of \cite{Kreft}. This procedure coalesces to  
\bea
a_{p/q}^{(\lambda)}(2j)&=& (-1)^{j+1}\sum_{k_1=0}^{q-2j}\sum_{k_2=0}^{k_1}\ldots\sum_{k_{j}=0}^{k_{j-1}} \alpha_{p/q}^{(\lambda)}(k_1+2j-2)\overline{\alpha}_{p/q}^{(\lambda)}(k_1+2j-2) \nonumber \\
&&\alpha_{p/q}^{(\lambda)}(k_2+2j-4)\overline{\alpha}_{p/q}^{(\lambda)}(k_2+2j-4)
\ldots \alpha_{p/q}^{(\lambda)}(k_{j})\overline{\alpha}_{p/q}^{(\lambda)}(k_{j}) \label{thealambda}
\eea
with for building blocks 
\be \nonumber\alpha_{p/q}^{(\lambda)}(k)\overline{\alpha}_{p/q}^{(\lambda)}(k),\ee
where
\be\nonumber\alpha_{p/q}^{(\lambda)}(k)=(\lambda/2)e^{-ik_y}\left(1-e^{{2  i\pi (k+1) p \over q}}\right)\ee 
and
\be\nonumber\overline{\alpha}_{p/q}^{(\lambda)}(k)=(\lambda/2)e^{ik_y}\left(1-({2/\lambda})^2 e^{-{2 i \pi (k+1) p \over q}}\right).\ee 
 When $\lambda\ne 2$,  even though $\overline{\alpha}_{p/q}^{(\lambda)}(k)$  is not anymore the complex conjugate of  $\alpha_{p/q}^{(\lambda)}(k)$, it is still true that $a_{p/q}^{(\lambda)}(2j)$ in \eqref{thealambda} is real. Indeed its imaginary part cancels because of
\be
\sum_{k=1}^{q} \rme^{\frac{2\rmi\pi k p}{q}j} = 0
\nonumber
\ee
when $q>j$, being a sum of $j$-th powers of $q$-th roots of unity\footnote{For example when $j=1$, 
\begin{align*}a_{p/q}^{(\lambda)}(2)= \sum_{k=0}^{q-2}\alpha_{p/q}^{(\lambda)}(k)\overline{\alpha}_{p/q}^{(\lambda)}(k)&=(\lambda/2)^2\sum_{k=0}^{q-2}\left(1-e^{{2 i \pi (k+1) p \over q}}+(2/\lambda)^2-(2/\lambda)^2e^{-{2 i \pi (k+1)  p \over q}}\right)\\&=(\lambda/2)^2\sum_{k=0}^{q-1}\left(1-e^{{2 i \pi (k+1) p \over q}}+(2/\lambda)^2-(2/\lambda)^2e^{-{2 i \pi (k+1) p \over q}}\right)\\&=(\lambda/2)^2\sum_{k=0}^{q-1}\left(1+(2/\lambda)^2\right)=q\left(1+(\lambda/2)^2\right).\end{align*}.}.

It follows that in the  $\lambda\ne 2$ almost Mathieu case  the $\pm s$ spectrum traces  and  their generating function   are  directly obtained  by replacing in  the Hofstadter $\pm s$ spectrum traces and   generating functions 
 \eqref{sumformulas} and \eqref{eq:K-eqs}  the $a_{p/q}(2j)$'s by the $a_{p/q}^{(\lambda)}(2j)$'s  and so the polynomial $b_{p/q}(z)$ by $b_{p/q}^{(\lambda)}(z)$. This is due to the purely algebraic  construction of these traces in terms of the $a_{p/q}(2j)$'s  via the roots of  \eqref{sss}  and therefore, in the $\lambda\ne 2$ case, the roots of   
\be\nonumber E^q b_{p/ q}^{(\lambda)}(1/E)=\pm s\ee 
with $0\le |s|\le 2\big(1+(\lambda/2)^q\big) $, since from \eqref{good} necessarily  $0\le| 2(\cos(q k_x)+({\lambda/ 2})^q\cos(q k_y))|\le 2\big(1+(\lambda/2)^q\big)$.

As stated in the introduction, one   wishes to obtain a closed formula  for the almost Mathieu quantum trace defined as
\be\label{mathieueq}
{\rm Tr}\; (H_{2\pi p/ q}^{(\lambda)})^n = \frac{1}{q}\int_{-\pi}^{\pi}\int_{-\pi}^{\pi} \frac{d k_x}{2\pi}\frac{d k_y}{2\pi} \sum_{r= 1}^{q} (E_r^{(\lambda)}(k_x,k_y) )^n,
\ee
where  
 the $E_r^{(\lambda)}(k_x,k_y)$'s are  the roots of \eqref{good}.
To do so, as in the Hofstadter case, one  aims at integrating  over  $s\in[-2(1+(\lambda/2)^q),2(1+(\lambda/2)^q)]$ the $\pm s$ spectrum trace     with a density of state $\rho^{(\lambda)}(s)$  
\be \int_{-2(1+(\lambda/2)^q)}^{2(1+(\lambda/2)^q)}{\rm Tr_{s}} ({H^{(\lambda)}_{2\pi p/q}})^n \rho^{(\lambda)}(s) ds= {\rm Tr} {(H^{(\lambda)}_{2\pi p/q})^n}  \nonumber \ee
 which has necessarily to be  such that       
\be \int_{-2(1+(\lambda/2)^q)}^{2(1+(\lambda/2)^q)}s^{2k} \rho^{(\lambda)}(s) ds={2k\choose k}\sum_{k_1=0}^{k}{k\choose k_1}^2(\lambda/2)^{2qk_1}\label{requiredlambda}\ee 
in analogy to~\eqref{required}.
A derivation of $\rho^{(\lambda)}(s)$  enforcing \eqref{requiredlambda} is given in the appendix. This density of states  amounts to  a $\lambda$-deformation of  $\rho(s)$
in \eqref{itis} for  the 2d lattice   spectrum 
 $s=2(\cos(q k_x)+(\lambda/2)^q\cos(q k_y))$.

Putting all the steps above together, namely  in \eqref{sumformulas} replacing both the $a_{p/q}(2j)$'s by the $a_{p/q}^{(\lambda)}(2j)$'s and  $s^{2k}$ by ${2k\choose k}\sum_{k_1=0}^{k}{k\choose k_1}^2(\lambda/2)^{2qk_1}$, the almost Mathieu operator quantum trace   \eqref{mathieueq} ends up as
\begin{align}\label{sumfinalement}
{\rm Tr} \;(H_{2\pi p/q}^{(\lambda)})^{n} =& \frac{n}{q} \sum_{k \geq 0} \sum_{\substack{\ell_1,\ell_2,\ldots,\ell_{\lfloor q/2 \rfloor} \geq 0 \\ \ell_1 + 2\ell_2 + \cdots + \lfloor q/2 \rfloor \ell_{\lfloor q/2 \rfloor} = n/2 - qk}}
\frac{\binom{\ell_1+\ell_2 + \cdots + \ell_{\lfloor q/2 \rfloor} + 2k}{\ell_1,\ell_2,\ldots,\ell_{\lfloor q/2 \rfloor},2k}}{\ell_1+\ell_2 + \cdots + \ell_{\lfloor q/2 \rfloor} + 2k} {2k\choose k}\sum_{k_1=0}^{k}{k\choose k_1}^2(\lambda/2)^{2qk_1}\\ \nonumber & \prod_{j = 1}^{\lfloor q/2 \rfloor} a_{p/q}^{(\lambda)}(2j)^{\ell_j},
\end{align}
which is a $\lambda$-deformation of the quantum Hofstadter trace (\ref{sumformula}).
 
 One also gets the generating function\footnote{\eqref{semi-final} can also be  obtained as in \cite{nous}  by considering lattice walks with asymmetric probability jumps on the horizontal axis versus the vertical axis in a ratio $\lambda$ (see the appendix).} that generalizes (\ref{eq:K-eq})
\be\label{semi-final}
\sum_{n \geq 0} Z_n^{(\lambda)}(e^{2i\pi p/q}) z^n =\Big( 1 - \frac{z(b_{p/ q}^{(\lambda)}(z))'}{qb_{p/ q}^{(\lambda)}(z)}\Big) \sum_{k \geq 0} \binom{2k}{k}\sum_{k_1=0}^k \binom{k}{k_1}^2 (\lambda/2)^{2q k_1}  \Big( \frac{z^q}{b_{p/ q}^{(\lambda)}(z)} \Big)^{2k},
\ee
where  $\sum_{k \geq 0} \binom{2k}{k}\sum_{k_1=0}^k \binom{k}{k_1}^2 (\lambda/2)^{2q k_1}  x^{k}$ can be viewed as  a $\lambda$-deformation  of the  complete elliptic integral of the first kind $\frac{2}{\pi} K(16x)=\sum_{k \geq 0} \binom{2k}{k}^2 x^k$.

Note finally that one can check that the Aubry duality  \cite{Aubry} 
\be\label{duality}{\rm Tr} \;(H_{2\pi p/q}^{(\lambda)})^{n}=({\lambda\over 2})^n{\rm Tr} \;(H_{2\pi p/q}^{(4/\lambda)})^{n},\ee 
holds,   as it should. This happens because the generalized Kreft coefficients \eqref{thealambda} obey  themselves the duality 
\be\label{dualitybis} a_{p/q}^{(\lambda)}(2j)=({\lambda\over 2})^{2j}\;a_{p/q}^{(4/\lambda)}(2j)\ee
which follows from
\be\nonumber{\rm Re}\left(\alpha_{p/q}^{(\lambda)}(k)\overline{\alpha}_{p/q}^{(\lambda)}(k)\right)= \Big({\lambda\over 2}\Big)^{2}\;{\rm Re}\left(\alpha_{p/q}^{(4/\lambda)}(k)\overline{\alpha}_{p/q}^{(4/\lambda)}(k)\right)\ee
(only the real part  is needed here since  $a_{p/q}^{(\lambda)}(2j)$ is real). The duality \eqref{duality} then follows from \eqref{sumfinalement} and the duality \eqref{dualitybis}
\begin{align*}
{\rm Tr} \;(H_{2\pi p/q}^{(\lambda)})^{n} &= \frac{n}{q} \sum_{k \geq 0} \sum_{\substack{\ell_1,\ell_2,\ldots,\ell_{\lfloor q/2 \rfloor} \geq 0 \\ \ell_1 + 2\ell_2 + \cdots + \lfloor q/2 \rfloor \ell_{\lfloor q/2 \rfloor} = n/2 - qk}}
\frac{{2k\choose k}\sum_{k_1=0}^{k}{k\choose k_1}^2(\lambda/2)^{2qk_1}\binom{\ell_1+\ell_2 + \cdots + \ell_{\lfloor q/2 \rfloor} + 2k}{\ell_1,\ell_2,\ldots,\ell_{\lfloor q/2 \rfloor},2k}}{\ell_1+\ell_2 + \cdots + \ell_{\lfloor q/2 \rfloor} + 2k} \\ \nonumber & \left(\lambda/2\right)^{\sum_{j=1}^{\lfloor q/2 \rfloor}2j l_j}\prod_{j = 1}^{\lfloor q/2 \rfloor} a_{p/q}^{(4/\lambda)}(2j)^{\ell_j}\\
&= \frac{n}{q} \sum_{k \geq 0} \sum_{\substack{\ell_1,\ell_2,\ldots,\ell_{\lfloor q/2 \rfloor} \geq 0 \\ \ell_1 + 2\ell_2 + \cdots + \lfloor q/2 \rfloor \ell_{\lfloor q/2 \rfloor} = n/2 - qk}}
\frac{{2k\choose k}\sum_{k_1=0}^{k}{k\choose k_1}^2(\lambda/2)^{2qk_1}\binom{\ell_1+\ell_2 + \cdots + \ell_{\lfloor q/2 \rfloor} + 2k}{\ell_1,\ell_2,\ldots,\ell_{\lfloor q/2 \rfloor},2k}}{\ell_1+\ell_2 + \cdots + \ell_{\lfloor q/2 \rfloor} + 2k}\\
\nonumber &
\left(\lambda/2\right)^{n-2qk}\prod_{j = 1}^{\lfloor q/2 \rfloor} a_{p/q}^{(4/\lambda)}(2j)^{\ell_j}\\
&= \left(\lambda/2\right)^{n}\frac{n}{q} \sum_{k \geq 0} \sum_{\substack{\ell_1,\ell_2,\ldots,\ell_{\lfloor q/2 \rfloor} \geq 0 \\ \ell_1 + 2\ell_2 + \cdots + \lfloor q/2 \rfloor \ell_{\lfloor q/2 \rfloor} = n/2 - qk}}
\frac{{2k\choose k}\sum_{k_1=0}^{k}{k\choose k_1}^2(2/\lambda)^{2qk-2qk_1}\binom{\ell_1+\ell_2 + \cdots + \ell_{\lfloor q/2 \rfloor} + 2k}{\ell_1,\ell_2,\ldots,\ell_{\lfloor q/2 \rfloor},2k}}{\ell_1+\ell_2 + \cdots + \ell_{\lfloor q/2 \rfloor} + 2k}\\
\nonumber &
\prod_{j = 1}^{\lfloor q/2 \rfloor} a_{p/q}^{(4/\lambda)}(2j)^{\ell_j}\\
\nonumber&
\stackrel{\text{$k^{\prime}_1=k-k_1$}}{=}\left(\lambda/2\right)^{n}\frac{n}{q} \sum_{k \geq 0} \sum_{\substack{\ell_1,\ell_2,\ldots,\ell_{\lfloor q/2 \rfloor} \geq 0 \\ \ell_1 + 2\ell_2 + \cdots + \lfloor q/2 \rfloor \ell_{\lfloor q/2 \rfloor} = n/2 - qk}}
\frac{{2k\choose k}\sum_{k^{\prime}_1=0}^{k}{k\choose k^{\prime}_1}^2(2/\lambda)^{2qk^{\prime}_1}\binom{\ell_1+\ell_2 + \cdots + \ell_{\lfloor q/2 \rfloor} + 2k}{\ell_1,\ell_2,\ldots,\ell_{\lfloor q/2 \rfloor},2k}}{\ell_1+\ell_2 + \cdots + \ell_{\lfloor q/2 \rfloor} + 2k}\\
\nonumber &
\prod_{j = 1}^{\lfloor q/2 \rfloor} a_{p/q}^{(4/\lambda)}(2j)^{\ell_j}=\left(\lambda/2\right)^n{\rm Tr} \;(H_{2\pi p/q}^{(4/\lambda)})^{n}
\end{align*}.

\section{Conclusion}
One has obtained the  $\lambda$-deformation of the quantum  trace  (\ref{sumformula}) in the form of (\ref{sumfinalement}). Both trace formulae have a similar structure, with clearly (\ref{sumfinalement}), the almost Mathieu case, reducing when $\lambda=2$ to (\ref{sumformula}), the Hofstadter case. Going back for a moment to random walks on a lattice, it would certainly be interesting to look at possible interpretations of (\ref{sumfinalement}) in the context of asymmetric paths with unequal  probabilities on the horizontal and vertical axis.

On the other hand, our results are directly relevant for problems related to  Calabi-Yau geometry. A recent work \cite{Hatsuda} indicates that there exists in the rational case a relation between the almost Mathieu operator and the relativistic Toda lattice. As pointed out in section 2 of \cite{Hatsuda}, there is an invariance under the modular double operation  exchanging $\gamma=2\pi p/q$ and  $\tilde{\gamma}=2\pi q/p$ (respectively $\hbar$ and $ \tilde{\hbar}$ in the notations of \cite{Hatsuda}) of the relativistic relative 2-body Toda Hamiltonian
\be H = \lambda(e^p+e^{-p})+e^x+e^{-x},\quad [x,p]=i\gamma\ee 
where,  in \cite{Hatsuda}, $\lambda$ is denoted as $R^2$.
The eigenvalues $E$ and $\tilde{E}$ corresponding  to $\gamma$ and $\tilde{\gamma}$ satisfy the polynomial identity $P_{p/q}(E)= P_{q/p}(\tilde{E})$ where  $P_{p/q}(E)$ is a    polynomial of degree $q$ (see $2.19$ in \cite{Hatsuda}). Now this polynomial is identical to the polynomial $E^q b_{p/q}^{(\lambda)}(1/E)$ introduced in \eqref{b_lambda},\eqref{surelambda}  which  encodes  the $k_x,k_y$-independent part of the determinant of the almost Mathieu Schrodinger  equation encapsulated in $m_{p/q}^{(\lambda)}(E,k_x,k_y)$. In the present work we have precisely obtained   in  (\ref{thealambda}) a  closed expression    for these polynomials in terms of the generalized Kreft coefficients.
 
It would certainly be rewarding to see if  \eqref{thealambda}   can bring any pertinent information related to the various qestions raised in \cite{Hatsuda}, in particlar regarding the quantum A-period for the  modulus of the underlying Calabi-Yau geometry. Finally, in the context of the Hofstadter model itself, the true role of the double modular transformation $\gamma\to \tilde{\gamma}$ remains to be elucidated.

\section{Acknowledgements}

S.W. was supported by the National Research Foundation of South Africa, grant number 96236.  S.O. would like to thank S. Nechaev for drawing his attention to \cite{Hatsuda}.

\section{Appendix}

\subsection{Kreft's coefficients construction:}
Following Kreft \cite{Kreft} we show how to get a closed expression for the polynomial \eqref{surelambda}, i.e. for the Kreft coefficients $a_{p/q}^{(\lambda)}(2j)$ in  $b_{p/ q}^{(\lambda)}(z)=-\sum_{j=0}^{[{q\over 2}]} a_{p/q}^{(\lambda)}(2j)z^{2j}$ as defined in \eqref{b_lambda}.

\subsubsection{\bf $\lambda= 2$:}

 One aims at transforming the matrix $m_{p/q}\left(E,k_x,k_y\right)$ in \eqref{matrixHof} into a tridiagonal one by an appropriate change of basis. 
First, let us do the change of basis
\begin{align*}
m_1=\begin{pmatrix}
1  & 0 & 0 & \cdots & 0 & 0 \\
0 & e^{ik_x}  & 0 & \cdots & 0 & 0 \\
0 &0 & e^{i 2k_x}  & \cdots & 0 & 0 \\
\vdots & \vdots & \vdots & \ddots & \vdots & \vdots \\
0 & 0 & 0  & \cdots & e^{i(q-2)k_x} & 0 \\
0 & 0 & 0  & \cdots & 0 & e^{i (q-1)k_x} \\
\end{pmatrix}
\end{align*}
so that $m_{p/q}\left(E,k_x,k_y\right)$ rewrites as 
\begin{align*}
m^{-1}_{1}m_{p/q}\left(E,k_x,k_y\right)m_1=
\begin{pmatrix}
2 \cos ({k_y})-E  & e^{ik_x} & 0 & \cdots & 0 & e^{-ik_x} \\
e^{-ik_x} & 2 \cos ({k_y}+\frac{2\pi p}{q})-E  & e^{ik_x} & \cdots & 0 & 0 \\
0 &e^{-ik_x} & ()  & \cdots & 0 & 0 \\
\vdots & \vdots & \vdots & \ddots & \vdots & \vdots \\
0 & 0 & 0  & \cdots & () & e^{ik_x} \\
e^{ik_x} & 0 & 0  & \cdots & e^{-ik_x} & 2 \cos ({k_y}+(q-1)\frac{2\pi p}{q})-E \\
\end{pmatrix}
\end{align*}
In this new basis the Schr\"odinger equation \eqref{eq} with ${\Phi}_{m+q}=e^{i q k_x}{\Phi}_m$ becomes 
\be\nonumber
e^{i k_x}\Phi_{m+1}'+e^{-i k_x}\Phi_{m-1}'+2\cos \left(k_y+\gamma m\right)\Phi_m'=E \Phi_m',\quad \Phi_{m+q}'=\Phi_m'
\ee

Then let us do a  second change of basis $m_2$
with matrix element
\begin{align*}
m_{jk}= \frac{1}{\sqrt{q}} (-1)^{p k}e^{\frac{2i\pi p}{q}\left(j k -k^2/2\right)}, \quad k, j \in\lbrace 0,1,\ldots,q-1\rbrace.
\end{align*} 
Putting together  $m_1$ and $m_2$ amounts to the change of basis  $m_1m_2$ with matrix element
$e^{i k k_x}m_{jk}= e^{i k k_x}\frac{1}{\sqrt{q}} (-1)^{p k}e^{\frac{2i\pi p}{q}\left(j k -k^2/2\right)}$ 
\begin{align}\label{newmatrix}
(m_1m_2)^{-1}m_{p/q}\left(E,k_x,k_y\right)m_1m_2=\begin{pmatrix}
-E  & \alpha_{p/q}(0) & 0 & \cdots & 0 & \overline{\alpha}_{p/q}(q-1) \\
\overline{\alpha}_{p/q}(0) & -E  & \alpha_{p/q}(1) & \cdots & 0 & 0 \\
0 &\overline{\alpha}_{p/q}(1) & -E  & \cdots & 0 & 0 \\
\vdots & \vdots & \vdots & \ddots & \vdots & \vdots \\
0 & 0 & 0  & \cdots & -E & \alpha_{p/q}({q-2}) \\
\alpha_{p/q}({q-1}) & 0 & 0  & \cdots & \overline{\alpha}_{p/q}({q-2}) & -E \\
\end{pmatrix}
\end{align} 
where $\alpha_{p/q}(k)=e^{-ik_y}-(-1)^{(p+1)}e^{\frac{2i\pi p}{q}(k+1/2)+ik_x}$, $\overline{\alpha}_{p/q}(k)$ is its complex conjugate and accordingly 
\be\label{aie}
\alpha_{p/q}(m){\tilde{\Phi}}_{m+1}+\overline{\alpha}_{p/q}(m-1){\tilde{\Phi}}_{m-1}=E{\tilde{\Phi}}_m,\quad {\tilde{\Phi}}_{m+q}={\tilde{\Phi}}_m
\ee

Both  corners $\alpha_{p/q}({q-1})$ and $\overline{\alpha}_{p/q}({q-1})$ in the matrix \eqref{newmatrix} can be cancelled if $e^{i(k_x+k_y)}=(-1)^{(p+1)}e^{-\frac{2i\pi p}{q}(q-1+1/2)}=(-1)^{(p+1)}e^{\frac{i\pi p}{q}}$, i.e. if $k_x + k_y = \pi(p+1) + \frac{\pi p}{q}$. The matrix is then tridiagonal  with a determinant in  \eqref{Chambers}  equal to  $(-1)^q E^q b_{p/q}(1/E)$, since  the  trigonometric part  vanishes as well: indeed 
\begin{align*}\cos(qk_x)+\cos(qk_y)&=2 \cos({q(k_x+k_y)\over 2})\cos({q(k_x-k_y)\over 2})\\ \nonumber &=2 \cos({\pi (p+q(p+1))\over 2})\cos({q(k_x-k_y)\over 2})\\ \nonumber &=0\end{align*}
due to $p+q(p+1)$ being always  odd since $p$ and $q$ are co-prime.

One gets the  tridiagonal matrix 
  \begin{align}\label{tridiag}
\begin{pmatrix}
-E  & \alpha_{p/q}(0) & 0 & \cdots & 0 & 0 \\
\overline{\alpha}_{p/q}(0) & -E  & \alpha_{p/q}(1) & \cdots & 0 & 0 \\
0 &\overline{\alpha}_{p/q}(1) & -E  & \cdots & 0 & 0 \\
\vdots & \vdots & \vdots & \ddots & \vdots & \vdots \\
0 & 0 & 0  & \cdots & -E & \alpha_{p/q}({q-2)} \\
0 & 0 & 0  & \cdots & \overline{\alpha}_{p/q}({q-2}) & -E \\
\end{pmatrix}
\end{align}
where $\alpha_{p/q}(k)$, using $e^{i(k_x+k_y)}=(-1)^{(p+1)}e^{\frac{i\pi p}{q}}$,  has simplified to $\alpha_{p/q}(k)=e^{-ik_y}(1-e^{\frac{2i\pi (k+1) p}{q}})$.
The matrix \eqref{tridiag}  does not depend anymore on $k_x$ nor $k_y$  and  its determinant satisfies a recursion (see  \cite{Kreft})
leading respectively to the Kreft polynomial and Kreft coefficients \eqref{sososimple} and \eqref{thea}.

\subsubsection{\bf $\lambda\ne 2$:}
One uses the same method as above to 
 find a closed expression for the  polynomial $b^{(\lambda)}_{p/q}(1/E)$, namely transform the matrix  $m_{p/q}^{(\lambda)}\left(E,k_x,k_y\right)$ in \eqref{matrixHoflambda} into a tridiagonal one.
To  do so  use the same change of basis  $m_1m_2$  as above so that
\begin{align}\label{onyest}
(m_1m_2)^{-1}m^{(\lambda)}_{p/q}\left(E,k_x,k_y,\right)m_1m_2=\begin{pmatrix}
-E  & \alpha_{p/q}^{(\lambda)}(0) & 0 & \cdots & 0 & \overline{\alpha}_{p/q}^{(\lambda)}({q-1}) \\
\overline{\alpha}_{p/q}^{(\lambda)}(0) & -E  & \alpha_{p/q}^{(\lambda)}(1) & \cdots & 0 & 0 \\
0 &\overline{\alpha}_{p/q}^{(\lambda)}(1) & -E  & \cdots & 0 & 0 \\
\vdots & \vdots & \vdots & \ddots & \vdots & \vdots \\
0 & 0 & 0  & \cdots & -E & \alpha_{p/q}^{(\lambda)}({q-2}) \\
\alpha_{p/q}^{(\lambda)}({q-1}) & 0 & 0  & \cdots & \overline{\alpha}_{p/q}^{(\lambda)}({q-2}) & -E \\
\end{pmatrix}
\end{align}
with $\alpha_{p/q}^{(\lambda)}(k)=({\lambda}/{2})e^{-ik_y}-(-1)^{p+1}e^{i\frac{2\pi p}{q}(k+1/2)+ik_x}$ and $ \overline{\alpha}_{p/q}^{(\lambda)}(k)$ its complex conjugate. Moreover, we have a resulting Schr\"odinger equation  identical to \eqref{aie}  provided that $\alpha_{p/q}(k)$ is replaced with $\alpha_{p/q}^{(\lambda)}(k)$.

Contrary to the Hofstadter case $\lambda=2$, both corners ${\alpha_{p/q}^{(\lambda)}({q-1})}$ and $\overline{\alpha}_{p/q}^{(\lambda)}({q-1})$  in the matrix \eqref{onyest} cannot   simultaneously vanish. One can still choose to have the lower left corner ${\alpha_{p/q}^{(\lambda)}({q-1})}$ to vanish: this
amounts to  $e^{i(k_x+k_y)}=(-1)^{p+1}({\lambda}/{2})e^{-i\frac{2\pi p}{q}(q-1+1/2)}=(-1)^{p+1}({\lambda}/{2})e^{\frac{i\pi p}{q}}$, which can only be achieved for a complex  $k_x+k_y$ (namely $k_x + k_y = -i \log(\lambda/2) + \frac{\pi p}{q} + \pi (p+1)$), so that    $e^{-i(k_x+k_y)}=(-1)^{p+1}({2}/{\lambda})e^{-\frac{i\pi p}{q}}$. The matrix  \eqref{onyest} then becomes
\begin{align}\label{almost}
\begin{pmatrix}
-E  & \alpha_{p/q}^{(\lambda)}(0) & 0 & \cdots & 0 & \overline{\alpha}_{p/q}^{(\lambda)}({q-1}) \\
\overline{\alpha}_{p/q}^{(\lambda)}(0) & -E  & \alpha_{p/q}^{(\lambda)}(1) & \cdots & 0 & 0 \\
0 &\overline{\alpha}_{p/q}^{(\lambda)}(1) & -E  & \cdots & 0 & 0 \\
\vdots & \vdots & \vdots & \ddots & \vdots & \vdots \\
0 & 0 & 0  & \cdots & -E & \alpha_{p/q}^{(\lambda)}({q-2}) \\
0 & 0 & 0  & \cdots & \overline{\alpha}_{p/q}^{(\lambda)}({q-2}) & -E \\
\end{pmatrix}
\end{align}
{where}  $\alpha_{p/q}^{(\lambda)}(k)$  and $\overline{\alpha}_{p/q}^{(\lambda)}(k)$ have simplified to
$\alpha_{p/q}^{(\lambda)}(k)=({\lambda}/{2})e^{-ik_y}(1-e^{\frac{2i\pi (k+1)p}{q}})$ and   $\overline{\alpha}_{p/q}^{(\lambda)}(k)=({\lambda}/{2})e^{ik_y}(1 - ({2}/{\lambda})^2 e^{- \frac{2 i\pi(k+1) p}{q} })$.   Note that  $
\overline{\alpha}_{p/q}^{(\lambda)}(k)$ is not anymore the complex conjugate of $
\alpha_{p/q}^{(\lambda)}(k)$.

By expanding the determinant of \eqref{almost}  with respect to  the elements of the first row
\small{\begin{align}
&\nonumber \begin{vmatrix}
-E  & \alpha_{p/q}^{(\lambda)}(0) & 0 & \cdots & 0 & \overline{\alpha}_{p/q}^{(\lambda)}({q-1}) \\
\overline{\alpha}_{p/q}^{(\lambda)}(0) & -E  & \alpha_{p/q}^{(\lambda)}(1) & \cdots & 0 & 0 \\
0 &\overline{\alpha}_{p/q}^{(\lambda)}(1) & -E  & \cdots & 0 & 0 \\
\vdots & \vdots & \vdots & \ddots & \vdots & \vdots \\
0 & 0 & 0  & \cdots & -E & \alpha_{p/q}^{(\lambda)}({q-2}) \\
0 & 0 & 0  & \cdots & \overline{\alpha}_{p/q}^{(\lambda)}({q-2}) & -E 
\end{vmatrix}= -E\begin{vmatrix}
 -E  & \alpha_{p/q}^{(\lambda)}(1) & \cdots & 0 & 0 \\
\overline{\alpha}_{p/q}^{(\lambda)}(1) & -E  & \cdots & 0 & 0 \\
 \vdots & \vdots & \ddots & \vdots & \vdots \\
 0 & 0  & \cdots & -E & \alpha_{p/q}^{(\lambda)}({q-2}) \\
 0 & 0  & \cdots & \overline{\alpha}_{p/q}^{(\lambda)}({q-2}) & -E
\end{vmatrix}\\&\nonumber -\alpha_{p/q}^{(\lambda)}(0)\overline{\alpha}_{p/q}^{(\lambda)}(0)\begin{vmatrix}
 -E  & \alpha_{p/q}^{(\lambda)}(2) & \cdots & 0 & 0 \\
\overline{\alpha}_{p/q}^{(\lambda)}(2) & -E  & \cdots & 0 & 0 \\
 \vdots & \vdots & \ddots & \vdots & \vdots \\
 0 & 0  & \cdots & -E & \alpha_{p/q}^{(\lambda)}({q-2}) \\
 0 & 0  & \cdots & \overline{\alpha}_{p/q}^{(\lambda)}({q-2}) & -E
\end{vmatrix}\\&+(-1)^{q+1}\overline{\alpha}_{p/q}^{(\lambda)}({q-1})\begin{vmatrix}
\overline{\alpha}_{p/q}^{(\lambda)}(0)  & -E & \alpha_{p/q}^{(\lambda)}(1)& \cdots & 0 & 0 \\
0 & \overline{\alpha}_{p/q}^{(\lambda)}(1) &-E  & \cdots & 0 & 0 \\
\vdots & \vdots & \vdots & \ddots & \vdots & \vdots \\
0 & 0 & 0  & \cdots & \overline{\alpha}_{p/q}^{(\lambda)}({q-3}) & -E \\
0 & 0 & 0  & \cdots & 0 & \overline{\alpha}_{p/q}^{(\lambda)}({q-2})
\end{vmatrix}\label{sharp}
\end{align}}
it is immediate to see that the part that depends on $k_x$ or  $k_y$ can only come from the last term of \eqref{sharp},  i.e. from the  upper right corner  $\overline{\alpha}_{p/q}^{(\lambda)}({q-1})$. Therefore to get the desired $k_x,k_y$-independent polynomial $E^q b^{(\lambda)}_{p/q}(1/E)$, all that is needed  is  the determinant of  the tridiagonal matrix
\begin{align}
\nonumber
\begin{pmatrix}
-E  & \alpha_{p/q}^{(\lambda)}(0) & 0 & \cdots & 0 & 0 \\
\overline{\alpha}_{p/q}^{(\lambda)}(0) & -E  & \alpha_{p/q}^{(\lambda)}(1) & \cdots & 0 & 0 \\
0 &\overline{\alpha}_{p/q}^{(\lambda)}(1) & -E  & \cdots & 0 & 0 \\
\vdots & \vdots & \vdots & \ddots & \vdots & \vdots \\
0 & 0 & 0  & \cdots & -E & \alpha_{p/q}^{(\lambda)}({q-2}) \\
0 & 0 & 0  & \cdots & \overline{\alpha}_{p/q}^{(\lambda)}({q-2}) & -E \\
\end{pmatrix}
\end{align}
which finally  yields   the    $\lambda\ne 2$  Kreft polynomial and Kreft coefficients \eqref{surelambda} and \eqref{thealambda}.

\subsection{Density of states $\rho^{(\lambda)}(s)$:}

\vspace{0.5cm}

To simplify the notations let us denote in this section $(\lambda/2)^q$ by $\tilde{\lambda}$. Knowing that
$$\int_{-4}^4 \rho(s) s^{2k}\,ds = \binom{2k}{k}^2$$
for $\rho(s) = (2\pi^2)^{-1} K(1-s^2/16)$, where $K$ denotes the elliptic integral, we would like to determine a function $\rho^{(\lambda)}(s)$ such that
$$\int_{-2(1+\tilde{\lambda})}^{2(1+\tilde{\lambda})} \rho^{(\lambda)}(s) s^{2k}\,ds = \binom{2k}{k} \sum_{k_1=0}^k \binom{k}{k_1}^2 \tilde{\lambda}^{2k_1}$$
as in \eqref{requiredlambda}.
The special case $\tilde{\lambda}=1$ clearly corresponds to the aforementioned formula.

\medskip

We  interpret the desired function $\rho^{(\lambda)}(s)$ as the density of a random variable with support $[-2(1+\tilde{\lambda}),2(1+\tilde{\lambda})]$ that is symmetric (so that the odd-order moments are $0$) and has $2k$-th moment
$$M_{2k} = \binom{2k}{k} \sum_{k_1=0}^k \binom{k}{k_1}^2 \tilde{\lambda}^{2k_1}.$$
The moment generating function associated with this random variable is
\begin{align*}
 \sum_{k=0}^{\infty} \frac{M_{2k}}{(2k)!} x^{2k} 
&= \sum_{k=0}^{\infty}\binom{2k}{k} \sum_{k_1=0}^k \binom{k}{k_1}^2 \tilde{\lambda}^{2k_1} \frac{x^{2k}}{(2k)!} \\
&= \sum_{k_1=0}^{\infty} \sum_{k=k_1}^{\infty} \binom{2k}{k} \binom{k}{k_1}^2 \tilde{\lambda}^{2k_1} \frac{x^{2k}}{(2k)!} \\
&= \sum_{k_1=0}^{\infty} \frac{\tilde{\lambda}^{2k_1}}{k_1!^2} \sum_{k=k_1}^{\infty} \frac{1}{(k-k_1)!^2}  x^{2k} \\
&= \sum_{k_1=0}^{\infty} \frac{\tilde{\lambda}^{2k_1}}{k_1!^2} \sum_{j=0}^{\infty} \frac{1}{j!^2}  x^{2(k_1+j)} \\
&= \sum_{k_1=0}^{\infty} \frac{\tilde{\lambda}^{2k_1}x^{2k_1}}{k_1!^2} \sum_{j=0}^{\infty} \frac{x^{2j}}{j!^2} \\
&= I(2\tilde{\lambda} x)I(2x),
\end{align*}
where $I$ denotes the $0$-th order modified Bessel function of the first kind. Thus the random variable whose density $\rho^{({\lambda})}(s)$ we would like to determine is the convolution of two random variables with moment generating functions $I(2\tilde{\lambda} x)$ and $I(2x)$ respectively.

\medskip

Now note that $I(2x)$ is exactly the moment generating function of an arcsine distribution on the interval $[-2,2]$:
$$\int_{-2}^2 \frac{e^{sx}}{\pi \sqrt{4-s^2}}\,ds = I(2x),$$
and likewise $I(2\tilde{\lambda} x)$ is the moment generating function of an arcsine distribution on the interval $[-2\tilde{\lambda},2\tilde{\lambda}]$:
$$\int_{-2\tilde{\lambda}}^{2\tilde{\lambda}} \frac{e^{sx}}{\pi \sqrt{4\tilde{\lambda}^2-s^2}}\,ds = I(2\tilde{\lambda} x).$$
Therefore $\rho^{({\lambda})}(s)$ must be the convolution of the two densities $h_1(s) = (\pi \sqrt{4-s^2})^{-1}$ ($s \in [-2,2]$, otherwise $h_1(s) = 0$) and $h_2(s) = (\pi \sqrt{4\tilde{\lambda}^2-s^2})^{-1}$ ($s \in [-2\tilde{\lambda},2\tilde{\lambda}]$, otherwise $h_2(s) = 0$), which is
$$\rho^{(\lambda)}(s) = \int_{-\infty}^{\infty} h_1(s-t) h_2(t)\,dt.$$
This reduces to several different cases:
\begin{itemize}
\item Case 1: $\tilde{\lambda} > 1$
\begin{enumerate}
\item if $s < -2-2\tilde{\lambda}$ or $s > 2+2\tilde{\lambda}$, then $\rho^{(\lambda)}(s) = 0$;
\item if $-2-2\tilde{\lambda} \leq s < 2-2\tilde{\lambda}$, then
$$\rho^{(\lambda)}(s) = \pi^{-2} \int_{-2\tilde{\lambda}}^{s+2} \frac{1}{\sqrt{(4-(s-t)^2)(4\tilde{\lambda}^2-t^2)}}\,dt;$$
\item if $2-2\tilde{\lambda} \leq s \leq 2\tilde{\lambda} -2$, then
$$\rho^{(\lambda)}(s) = \pi^{-2} \int_{s-2}^{s+2} \frac{1}{\sqrt{(4-(s-t)^2)(4\tilde{\lambda}^2-t^2)}}\,dt;$$
\item if $2\tilde{\lambda}-2 < s \leq 2\tilde{\lambda}+2$, then
$$\rho^{(\lambda)}(s) = \pi^{-2} \int_{s-2}^{2\tilde{\lambda}} \frac{1}{\sqrt{(4-(s-t)^2)(4\tilde{\lambda}^2-t^2)}}\,dt.$$
\end{enumerate}
\item Case 2: $0 < \tilde{\lambda} < 1$
\begin{enumerate}
\item if $s < -2-2\tilde{\lambda}$ or $s > 2+2\tilde{\lambda}$, then $\rho^{(\lambda)}(s) = 0$;
\item if $-2-2\tilde{\lambda} \leq s < 2\tilde{\lambda} -2$, then
$$\rho^{(\lambda)}(s) = \pi^{-2} \int_{-2\tilde{\lambda}}^{s+2} \frac{1}{\sqrt{(4-(s-t)^2)(4\tilde{\lambda}^2-t^2)}}\,dt;$$
\item if $2\tilde{\lambda}-2 \leq s \leq 2-2\tilde{\lambda}$, then
$$\rho^{(\lambda)}(s) = \pi^{-2} \int_{-2\tilde{\lambda}}^{2\tilde{\lambda}} \frac{1}{\sqrt{(4-(s-t)^2)(4\tilde{\lambda}^2-t^2)}}\,dt;$$
\item if $2-2\tilde{\lambda} < s \leq 2\tilde{\lambda} + 2$, then
$$\rho^{(\lambda)}(s) = \pi^{-2} \int_{s-2}^{2\tilde{\lambda}} \frac{1}{\sqrt{(4-(s-t)^2)(4\tilde{\lambda}^2-t^2)}}\,dt.$$
\end{enumerate}
\end{itemize}

\subsection{Eq.~\eqref{semi-final} can be directly obtained as in \cite{nous}:}
In order to reproduce  from the lattice walks formulation in \cite{nous} the results obtained above for $\lambda\ne  2$,  all that is needed is to set in  the generating function $A(x_1,x_2,y_1,y_2,\scq)$ (defined in eq.(10) of \cite{nous}) $x_1\rightarrow zx$, $x_2\rightarrow z/x$, $y_1\rightarrow (\lambda/2)zy$ and  $y_2\rightarrow (\lambda/2)z/y$ and look as in \cite{nous} at the coefficients with vanishing  exponents in $x$ and $y$  (i.e. $m_1-m_2=0$ and $l_1-l_2=0$) and  exponent  $n$ in $z$ (i. e., $m_1+m_2+l_1+l_2=n$) 
\begin{align}\label{generating function A}
\left[x^0y^0\right]A\left(zx,z/x,(\lambda/2)zy,(\lambda/2)z/y\right)=\left[x^0y^0\right]\frac{\begin{vmatrix}
1  & -z/x & 0 & \cdots & 0 & -zx \\
1 & c^{(\lambda)}_1  & -z/x & \cdots & 0 & 0 \\
1 &-zx & c^{(\lambda)}_2  & \cdots & 0 & 0 \\
\vdots & \vdots & \vdots & \ddots & \vdots & \vdots \\
1 & 0 & 0  & \cdots & c^{(\lambda)}_{q-2} & -z/x \\
1 & 0 & 0  & \cdots & -zx & c^{(\lambda)}_{q-1} \\
\end{vmatrix}}{\begin{vmatrix}
c^{(\lambda)}_0  & -z/x & 0 & \cdots & 0 & -zx \\
-zx & c^{(\lambda)}_1  & -z/x & \cdots & 0 & 0 \\
0 & -zx & c^{(\lambda)}_2  & \cdots & 0 & 0 \\
\vdots & \vdots & \vdots & \ddots & \vdots & \vdots \\
0 & 0 & 0  & \cdots & c^{(\lambda)}_{q-2} & -z/x \\
-z/x & 0 & 0  & \cdots & -zx & c^{(\lambda)}_{q-1} \\
\end{vmatrix}}
\intertext{where $c^{(\lambda)}_k=1-(\lambda/2)z(\scq^k y+\scq^{-k}y^{-1})$ with $\scq=e^{2i\pi p/q}$. }\nonumber
\end{align}
As  in  \cite{nous}   the denominator has  the form
\begin{align*}
\Delta(zx,z/x,(\lambda/2)zy,(\lambda/2)z/y)=1-z^q\left(x^q+x^{-q}+(\lambda/2)^q(y^q+y^{-q})\right)+V(z^2,(\lambda/2)^2z^2).
\end{align*}
 For $x=y=1$, we get \begin{align*}\Delta(z,z,(\lambda/2)z,(\lambda/2)z)=1-z^q\left(2+2(\lambda/2)^q\right)+V(z^2,(\lambda/2)^2z^2) 
\end{align*} 
so that
\begin{multline*}
\Delta(zx,z/x,(\lambda/2)zy,(\lambda/2)z/y) \\
 = \Delta(z,z,(\lambda/2)z,(\lambda/2)z)+z^q(2+2(\lambda/2)^q)-z^q\left(x^q+x^{-q}+(\lambda/2)^q(y^q+y^{-q})\right).
\end{multline*}
On the other hand recall that both the polynomial $b^{(\lambda)}_{p/q}(z)=-\sum^{[\frac{q}{2}]}_{j=0}a^{(\lambda)}_{p/q}(2j)z^{2j}$ defined in \eqref{b_lambda} and the matrix $m^{(\lambda)}_{p/q}(e,0,0)$  defined in \eqref{matrixHoflambda} are related by \eqref{surelambda}.  
Since also
\begin{align*}\Delta\left(z,z,(\lambda/2)z,(\lambda/2)z\right)=(-z)^q{\rm det}\left(m^{(\lambda)}_{p/q}(1/z,0,0)\right)
\end{align*}
it follows that
\begin{align*}\Delta\left(z,z,(\lambda/2)z,(\lambda/2)z\right)=b^{(\lambda)}_{p/q}(z)-z^q(2+2(\lambda/2)^q)
\end{align*}
and finally
\begin{align*}
\Delta\left(zx,z/x,(\lambda/2)zy,(\lambda/2)z/y\right)=b^{(\lambda)}_{p/q}(z)-z^q\left(x^q+x^{-q}+(\lambda/2)^q(y^q+y^{-q})\right).
\end{align*}
Accordingly, and as shown in the appendix of \cite{nous} (see the proof of (18))  the numerator  at order $x^0y^0$  ends up being expressed only in terms of $b^{(\lambda)}_{p/q}(z)$
\begin{align*}
\left[x^0y^0\right]\begin{vmatrix}
1  & -z/x & 0 & \cdots & 0 & -zx \\
1 & c^{(\lambda)}_1  & -z/x & \cdots & 0 & 0 \\
1 &-zx & c^{(\lambda)}_2  & \cdots & 0 & 0 \\
\vdots & \vdots & \vdots & \ddots & \vdots & \vdots \\
1 & 0 & 0  & \cdots & c^{(\lambda)}_{q-2} & -z/x \\
1 & 0 & 0  & \cdots & -zx & c^{(\lambda)}_{q-1} \\
\end{vmatrix}=b^{(\lambda)}_{p/q}(z)-\frac{z}{q}b^{(\lambda)}_{p/q}(z)^{\prime}
\end{align*} 
Finally \eqref{generating function A} rewrites as
\begin{align*}
\left[x^0y^0\right]A\left(zx,z/x,(\lambda/2)zy,(\lambda/2)z/y\right)
&=[x^0y^0]\frac{b^{(\lambda)}_{p/q}(z)-\frac{z}{q}b^{(\lambda)}_{p/q}(z)^{\prime}}{b^{(\lambda)}_{p/q}(z)-z^q\left(x^q+x^{-q}+(\lambda/2)^q(y^q+y^{-q})\right)}\\
&=\left(1-\frac{zb^{(\lambda)}_{p/q}(z)^{\prime}}{qb^{(\lambda)}_{p/q}(z)}\right)\sum_{k\geq 0}\left(\frac{z^q}{b^{(\lambda)}_{p/q}(z)}\right)^k[x^0y^0]\left(x^q+x^{-q}+(\lambda/2)^q(y^q+y^{-q})\right)^k
\end{align*}
which yields  the generating function \eqref{semi-final} for the traces $ {\rm Tr}\left(H^{(\lambda)}_{2\pi p/q}\right)^n$: we have
\begin{align*}
\left[x^0y^0\right]\left(x^q+x^{-q}+(\lambda/2)^q(y^q+y^{-q})\right)^k&=\left[x^0y^0\right]\sum_{k_1+k_2+k_3+k_4=k}\binom{k}{k_1,k_2,k_3,k_4}\left(\lambda/2\right)^{q(k_1+k_2)}x^{q(k_3-k_4)}y^{q(k_1-k_2)}\\
&=\sum_{\substack{ k_1+k_3=k/2 \\k\;{\rm even}}}\binom{k}{k_1,k_1,k_3,k_3}\left(\lambda/2\right)^{2qk_1}\\
&=\binom{k}{k/2}\sum_{k_1=0}^{k/2}\binom{k/2}{k_1}^2\left(\lambda/2\right)^{2qk_1}
\end{align*}
The coefficient at order  $x^0y^0$ of $\left(x^q+x^{-q}+(\lambda/2)^q(y^q+y^{-q})\right)^k$ is $\binom{k}{k/2}\sum_{k_1=0}^{k/2}\binom{k/2}{k_1}^2\left(\lambda/2\right)^{2qk_1}$ if $k$ is even and $0$ otherwise,  so that finally
\begin{align*}
\sum_{n\geq 0, n\;{\rm even}}{\rm Tr} {(H^{(\lambda)}_{2\pi p/q})^n} z^n 
=\left(1-\frac{zb^{(\lambda)}_{p/q}(z)^{\prime}}{qb^{(\lambda)}_{p/q}(z)}\right)\sum_{k\geq 0}\left(\frac{z^q}{b^{(\lambda)}_{p/q}(z)}\right)^{2k}\binom{2k}{k}\sum_{k_1=0}^{k}\binom{k}{k_1}^2\left(\lambda/2\right)^{2qk_1}.
\end{align*}

\end{document}